\documentclass{article}
\usepackage{graphicx}
\author{P. Provero}
\title{Gene networks from DNA microarray data: centrality and lethality}
\date{}
\begin{document}
\maketitle
\noindent
{\it Dipartimento di Scienze e Tecnologie Avanzate, Universit\`a del
Piemonte Orientale, and INFN, gruppo collegato di Alessandria, 
Italy}
\vskip0.1cm\noindent
\begin{center}
and
\end{center}
\vskip0.1cm\noindent
{\it Dipartimento di Fisica
Teorica, Universit\`a di Torino, Via P. Giuria 1, 10125 Torino,
Italy}
\vskip0.5cm
e-mail: provero@to.infn.it
\vskip0.5cm
\begin{abstract}
\noindent
We construct a gene network based on expression
data from DNA microarray experiments, by establishing a link between
two genes whenever the Pearson's correlation coefficient between their
expression profiles is higher than a certain cutoff. The resulting connectivity
distribution is compatible with a power-law decay with exponent
$\gamma\sim 1$, corrected by an exponential cutoff at large
connectivity. The biological relevance of such network is
demonstrated by showing that there is a strong statistical correlation
between high connectivity number and lethality: in close analogy to
what happens for protein interaction networks, essential genes 
are strongly overerpresented among the
hubs of the network, that is the genes connected to many other genes. 
\end{abstract}
DNA microarray experiments are one of the most powerful tools for
studying interactions between genes on the scale of the whole
genome. It is widely believed that a huge amount of biologically
relevant information is encoded in the results of such experiments,
and that new analytical methods need to be developed to extract it.
\par
In this work we propose to analyse the expression data obtained in
microarray experiments by constructing a network of coregulated genes: 
the genes are the nodes of the network, and a link is established
between two nodes 
whenever they are similarly expressed across many experimental
conditions. On one hand, we show that such network, like many other
known networks of self-organizing origin, 
shows a connectivity distribution that decays with a power law
corrected, for large values of the connectivity, by an exponential
cutoff \cite{amaral:2000,albert:2001}.
On the other hand, we show that the network
encodes biologically relevant information in its topology: exactly as
in the case of the protein interaction network in yeast
\cite{jeong:2001}, 
centrality is strongly
correlated to lethality. In other words, among the genes that have the
highest connectivity in the network, essential genes, whose deletion
produces an inviable mutant, are strongly overrepresented.
\par
We will work on yeast ({\it S. cerevisiae}), 
and use the expression data made publicly
available by the authors of Ref.\cite{spellman:1998} 
(and that include also the data
obtained by the authors of Ref.\cite{cho:1998}), who performed a series of
microarray experiments with the goal of identifying cell-cycle
regulated genes. The data consist in the expression profiles for
virtually all of the $\sim 6200$ yeast genes across a total of 77
timepoints.
\par 
The network is constructed by the following procedure:
\begin{enumerate}
\item
To each gene we associate its expression profile defined as a string
of 77 real numbers, representing as it is customary the $log_2$ of the
ratio between expression (quantity of mRNA) at the given time-point
and a reference value of the expression. The data come from different
experiments and have been processed by the authors of
Ref.\cite{spellman:1998}, to which
we refer for details, so as to be comparable to each other across the
various experiments. 
\item
Missing values are replaced with the average expression over the available
timepoints. To prevent this manipulation from having a sizable effect
on the construction of the network, we retain only those
genes for which at least 70 timepoints out of 77 are available. We are thus left
with 5293 genes as the nodes of our network.
\item
We compute the Pearson's correlation coefficient $r$ for all pairs of
nodes in the network.
\item
We establish a link between two nodes whenever $r$ is larger than a
certain cutoff $C$.  
\end{enumerate}
The only free parameter in the procedure is the cutoff $C$. A possible
way of choosing it is to compare the probability distribution of $r$
for the actual data to the same distribution after the data have
been randomized by shuffling the expression values of each gene. For
the randomized data, no pair of genes shows a value of $r$ greater than
$0.67$: therefore by choosing $C=0.67$ the links we create can
be considered of biological origin. A similar procedure was used in
Ref.\cite{farkas:2001}, where a network was constructed by
establishing a link between two genes whenever the effects of their
deletion on the expression of the rest of the genome were linearly correlated.
\par
With this choice of the cutoff $C$, 17643 links are established between
the genes. Defining the connectivity $k$ of a node as the number of
links departing from it, we have an average connectivity $\langle
k\rangle\sim 6.67$. Defining  $N(k)$ as the number of nodes with
connectivity $k$, we see that $N(k)$ shows a 
long tail that reaches up to $k=173$: the shape of
the distribution is compatible with a 
power law decay with exponential cutoff:
\begin{equation}
N(k)=a\ k^{-\gamma}\exp\left(-\frac{k}{k_c}\right)
\label{pk}
\end{equation}
This is shown in Figs. 1-3: Fig. 1 shows $N(k)$ in logarithmic scale
as a function of
$k$ (noise in the data has been reduced by logarithmic binning),
showing that the decay of $N(k)$ is
slower than exponential for small to moderate values of $k$. 
Fig. 2 shows the same data with logarithmic
scale on the $k$ axis too, and demonstrates that the decay is faster,
at large $k$,
than the pure power law characteristic of scale-free networks. Finally
Fig. 3 shows the data after correction with the exponential cutoff,
with $k_c\sim 38$. The slope of the straight
line is $\gamma=0.95$. An analysis of the cumulative distribution
confirms these results. Interestingly, a recent study of the
transcriptional regulation network in yeast also revealed a scale-free
network with $\gamma\sim 1$ \cite{guelzim:2002}.
\begin{figure}
\centering
{\includegraphics[width=12.cm]{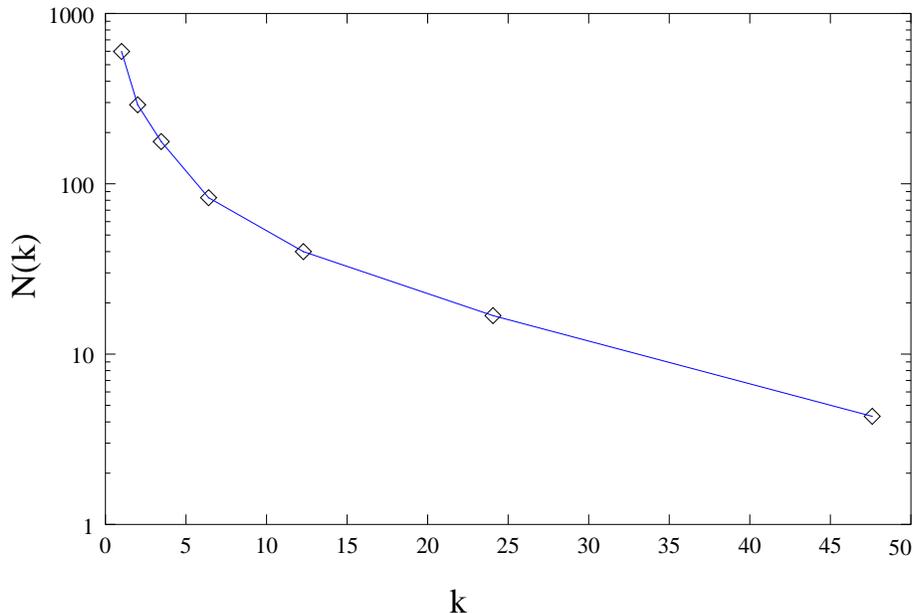}}
\caption{Linear-log plot of $N(k)$, the number of nodes with
connectivity $k$, showing that the decay is slower than exponential.} 
\end{figure}
\begin{figure}
\centering
{\includegraphics[width=12.cm]{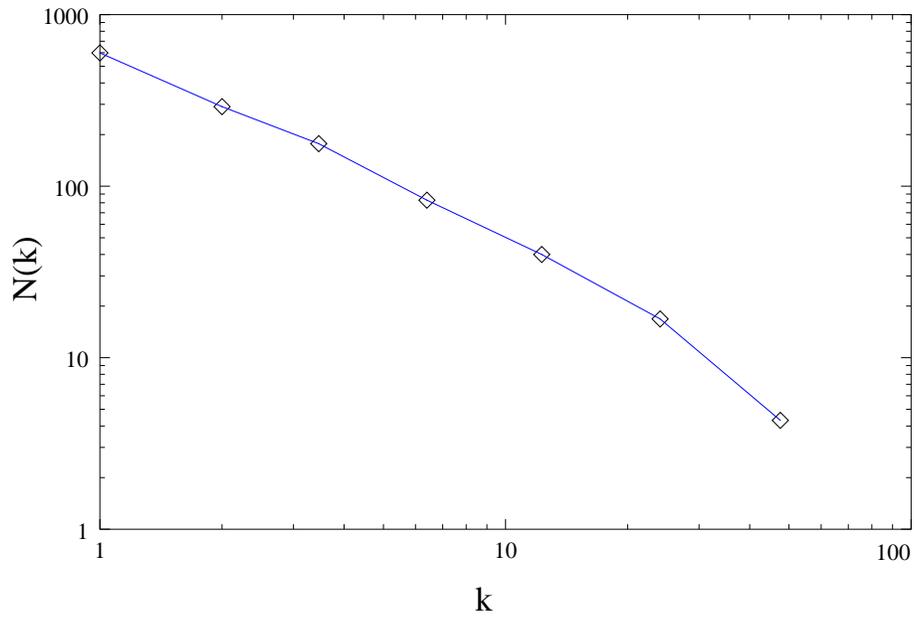}}
\caption{Log-log plot of $N(k)$, 
showing a power-like decay with an exponential cutoff at large distances.} 
\end{figure}
\begin{figure}
\centering
{\includegraphics[width=12.cm]{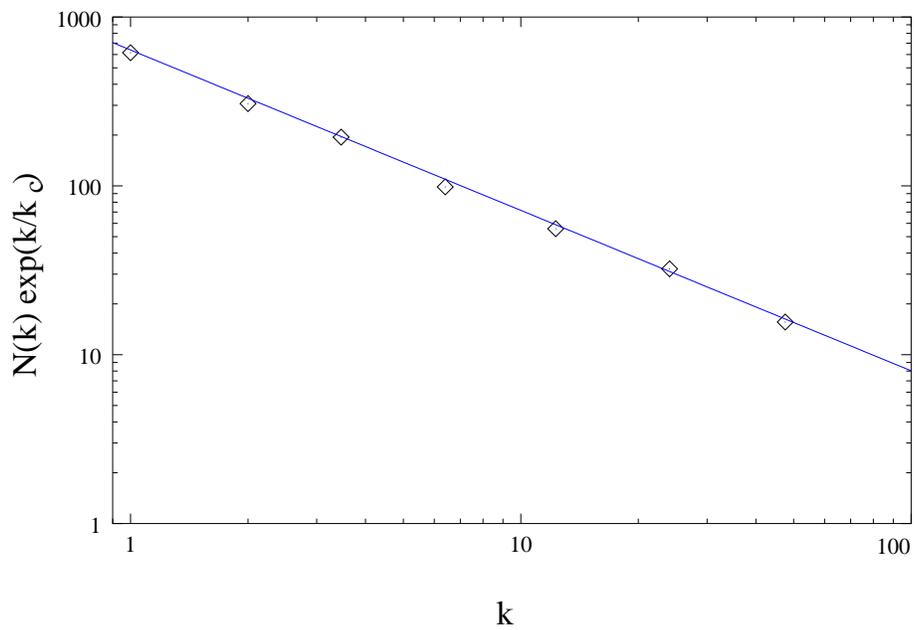}}
\caption{Log-log plot of $N(k)$ after correction with the exponential
cutoff at $k_c\sim 37$. The slope of the straight line is $\sim 0.95$.} 
\end{figure}
\par
In this paper, we are mainly concerned with establishing the
biological relevance of this network, independently of any of its
graph-theoretic features. This we will do by showing that the nodes in
the network with high connectivity are more likely to
be essential genes,  whose deletion produces an inviable
mutant. 
\par
A list of essential yeast genes is publicly available from the {\it
Saccharomyces Genome Deletion Project} \cite{winzeler:1999}, and
comprises 1104 genes, corresponding to 18.7\% of the genes deleted in
the project. Of the 5293 genes in our network, 964 (18.2\%) are
included in the list. Fig. 4 shows the fraction $f(k_{min})$ of essential genes
among the genes having connectivity $k_{min}$ or more, as a function of
$k_{min}$: it grows from $0.182$ at $k_{min}=0$ (by definition) to $1$ for the
the 6 most connected genes ($k\ge 155$). 
\par
The figure shows that essential genes are more and more
overrepresented as the minimum connectivity $k_{min}$ is increased.
To evaluate the statistical significance of such overrepresentation,
suppose that the number of genes with connectivity $k\ge k_{min}$ is
$n$, and that among these $m$ are essential. Then one can
evaluate the probability $P(N,M;n,m)$ that, out of
$n$ randomly chosen nodes out of a set of $N$, $m$ or more are
essential genes, when the total number of essential genes is $M$.  
This probability can be computed as the right tail of the
appropriate hypergeometric distribution, and reaches very small values:
for example the fraction of essential genes reaches 50\% for
$k_{min}=37$, with $m= 127$ essential genes out of $n=251$ nodes, and the
probability $P(N=5293,M=964;n=251,m=127)$ 
is about $4\cdot 10^{-33}$.
\par
In conclusion, we have built a gene network based on expression data
obtained with DNA microarray experiment, by joining genes showing
similar expression profiles. The resulting network shows a power law
decay of the connectivity distribution with
an exponential cutoff, and exponent $\gamma\sim 1$. Its biological
relevance is proved by the strong statistical correlation between
centrality and lethality. 
\begin{figure}
\centering
{\includegraphics[width=12.cm]{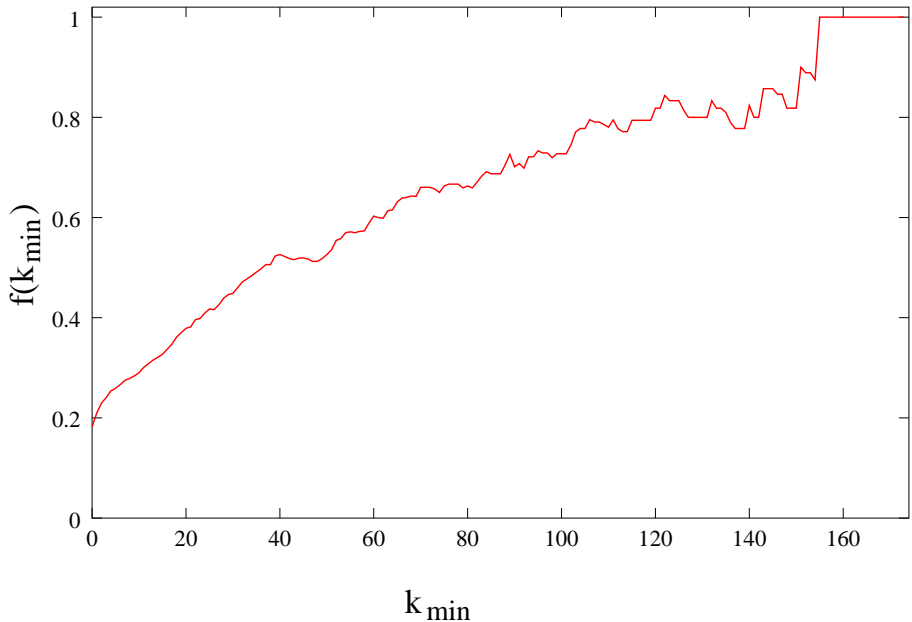}}
\caption{The fraction $f(k_{min})$ of essential genes among the ones
with connectivity $k\ge k_{min}$.}  
\end{figure}
\vskip1.cm\noindent
{\bf Acknowledgements.} I would like to thank M. Caselle and
C. Herrmann for many useful conversations and suggestions. I am also
grateful to M. Barth\'el\'emy for many useful comments about a
previous version of this manuscript.

\end{document}